\documentclass[11pt,a4paper]{article}
\usepackage[latin1]{inputenc}
\usepackage[T1]{fontenc}
\usepackage[english]{babel}
\usepackage{amssymb}
\usepackage{amsfonts}
\usepackage[centertags,intlimits]{amsmath}
\usepackage{theorem}
\usepackage[final]{epsfig}
\usepackage{enumerate}
\usepackage{graphicx}
\usepackage{showlabels}
\usepackage{cite}

\input{tcilatex}

\begin{document}

\title{\center{ \Large \textbf{$q$-deformed conformal correlation functions}}%
}
\author{L. Mesref\thanks{%
Email: l\_mesref@hotmail.com}}
\date{Département d'électrotechnique, \ faculté de Génie électrique\\
Université des Sciences et de la Technologie d'Oran \\
U.S.T.O. "Mohamed Boudiaf", Oran, Algeria}
\maketitle

\begin{abstract}
In this paper, we compute the general structure of two and three point
functions in field theories that are assumed to possess an invariance under
a quantum deformation of SO(4,2). The computation is elaborated in order to
fit the Hopf algebra structures.
\end{abstract}

\newpage

\section{Introduction}

Over the last thirty years, the subject of quantum groups \cite{manin,
drinfeld, jimbo, woronowicz} has grown into a full-fledged research topic. A
huge amount of literature has appeared. One can mention $q$-harmonic
analysis and $q$-special functions \cite{lustig}, conformal field theories 
\cite{alvarez, moore}, in the vertex and spin models \cite{vega, pasquier},
anyons \cite{lerda, roy, ubriaco}, in quantum optics \cite{buzek}, in the
loop approach of quantum gravity \cite{major} in \textquotedblleft fuzzy
physics\textquotedblright\ \cite{madore1} and quantum gauge theories \cite%
{bernard, mesref}. \newline
The deformation of two-dimensional conformally invariant field theory was
introduced in Ref. \cite{bernard2} and the properties of the correlation
functions were determined. The coherent states operators and the invariant
correlation functions and their quantum group counterparts, both for $%
sl\left( 2\right) $ and $SU_{3}$, were studied in \cite{furlan}. \newline
The $q$-deformation of $D=4$ conformal algebra was first introduced in \cite%
{dobrev}. It was also studied, and additionally its contraction to deformed
Poincare algebra given in Ref. \cite{luki}. The reality conditions of
deformed $SO\left( 6,\mathbb{C}\right) $ were discussed in \cite{lukierski}. 
\newline
In this paper, we propose a four-dimensional conformal field theory based on
the quantum universal enveloping algebra $U_{q}\left( so\left( 4,2\right)
\right) $ \cite{dobrev, luki}. In the course of these investigations we rely
on Dobrev's approach \cite{dobrev3}. According to this procedure one first
needs $q$-difference realizations of the representations in terms of
functions of non-commuting variables. These variables generate a flag
manifold of the matrix quantum group $SL_{q}\left( 4\right) $ which is in
duality with $U_{q}\left( sl\left( 4;\mathbb{C}\right) \right) $. \newline
This paper is organized as follows. In Section 2 we recall the Cartan-Weyl
basis of the quantum Lie algebra $U_{q}\left( sl\left( 4;\mathbb{C}\right)
\right) $. In Section 3, we compute the $q$-deformed two- and three-point
conformal correlation functions.

\section{Cartan-Weyl basis for $U_{q}\left( sl\left( 4;\mathbb{C}\right)
\right) $}

\setcounter{equation}{0} The positive energy irreducible representations of $%
so\left( 4,2\right) $ are labelled by the lowest value of the energy $E_{0}$%
, the spin $s_{0}=j_{1}+j_{2}$ and by the helicity $h_{0}=j_{1}-j_{2}$, and
these are eigenvalues of a Cartan subalgebra $\mathcal{H}$ of $so\left(
4,2\right) $. We shall label the representations of $U_{q}\left( so\left(
4,2\right) \right) $ in the same way and thus we shall take for $U_{q}\left(
so\left( 4,2\right) \right) $ and its complexification $U_{q}\left( so\left(
6,\mathbb{C}\right) \right) $ the same Cartan subalgebra. We recall that the 
$q$-deformation $U_{q}\left( so\left( 6,\mathbb{C}\right) \right) $ is
defined \cite{drinfeld,jimbo} as the associative algebra over $\mathbb{C}$
with Chevalley generators $X_{j}^{\pm }$, $H_{j}$, $j=1,2,3.$

The Cartan-Chevalley basis of $U_{q}\left( sl\left( 4,\mathbb{C}\right)
\right) $ is given by the formulae:

\bigskip

\begin{eqnarray}
\left[ H_{j},H_{k}\right] &=&0\qquad \left[ H_{j},X_{k}^{\pm }\right] =\pm
\,a_{jk}\,X_{k}^{\pm }\qquad \qquad \qquad \qquad \qquad  \notag \\
\left[ X_{j}^{+},X_{k}^{-}\right] &=&\delta _{jk}\frac{q^{H_{j}}-q^{-H_{j}}}{%
q-q^{-1}}=\delta _{jk}\left[ H_{j}\right] _{q}
\end{eqnarray}

\bigskip

and the $q$-analogue of the Serre relations

\bigskip

\begin{equation}
\left( X_{j}^{\pm }\right) ^{2}X_{k}^{\pm }-\left[ 2\right] _{q}X_{j}^{\pm
}X_{k}^{\pm }X_{j}^{\pm }+X_{k}^{\pm }\left( X_{j}^{\pm }\right) ^{2}=0,
\end{equation}

\bigskip

where $\left( jk\right) =\left( 12\right) ,\left( 21\right) ,\left(
23\right) ,\left( 32\right) $ and $\left( a_{jk}\right) $ is the Cartan
matrix of $so\left( 6,\mathbb{C}\right) $ given by $\left( a_{jk}\right)
=2\left( \alpha _{j},\alpha _{k}\right) /\left( \alpha _{j},\alpha
_{j}\right) $; $\alpha _{1},\alpha _{2},\alpha _{3}$ are the simple roots of
length 2 and the non-zero product between the simple roots are: $\left(
\alpha _{1},\alpha _{2}\right) =\left( \alpha _{2},\alpha _{3}\right) =-1$ .
The quantum number is defined as $\left[ m\right] _{q}=\frac{q^{m}-q^{-m}}{%
q-q^{-1}}$.

\bigskip Explicitly the Cartan matrix is given by:

\bigskip

\begin{equation}
\left( a_{jk}\right) =\left( 
\begin{array}{ccc}
2 & -1 & 0 \\ 
-1 & 2 & -1 \\ 
0 & -1 & 2%
\end{array}
\right) .
\end{equation}

\bigskip

The elements $H_{j}$ span the Cartan subalgebra $\mathcal{H}$ while the
elements $X_{j}^{\pm }$ generate the subalgebra $\mathcal{G}^{\pm }$ in the
standard decomposition $\mathcal{G}\equiv so\left( 6,\mathbb{C}\right) =%
\mathcal{G}^{+}\oplus \mathcal{H}\oplus \mathcal{G}^{-}$. In particular, the
Cartan-Weyl generators for the non-simple roots are given by \cite{dobrev}:

\bigskip

\begin{eqnarray}
X_{jk}^{\pm } &=&\pm q^{\mp 1/2}\left( q^{1/2}X_{j}^{\pm }X_{k}^{\pm
}-q^{-1/2}X_{k}^{\pm }X_{j}^{\pm }\right) \qquad \left( jk\right) =\left(
12\right) ,\left( 23\right)  \notag \\
X_{13}^{\pm } &=&\pm q^{\mp 1/2}\left( q^{1/2}X_{1}^{\pm }X_{23}^{\pm
}-q^{-1/2}X_{23}^{\pm }X_{1}^{\pm }\right)  \notag \\
&=&\pm q^{\mp 1/2}\left( q^{1/2}X_{12}^{\pm }X_{3}^{\pm }-q^{-1/2}X_{3}^{\pm
}X_{12}^{\pm }\right) .
\end{eqnarray}

\bigskip

All other commutation relations follow from these definitions:\quad 
\begin{eqnarray}
\left[ X_{a}^{+},X_{ab}^{-}\right] &=&-q^{H_{a}}\,X_{a+1b}^{-}\qquad \left[
X_{b}^{+},X_{ab}^{-}\right] =X_{ab-1}^{-}\,q^{-H_{b}}\quad 1\leq a<b\leq 3 
\notag \\
\left[ X_{a}^{-},X_{ab}^{+}\right] &=&X_{a+1b}^{+}\,q^{-H_{a}}\qquad \left[
X_{b}^{-},X_{ab}^{+}\right] =-q^{H_{b}}\,X_{ab-1}^{+}\quad 1\leq a<b\leq 3 
\notag \\
X_{a}^{\pm }X_{ab}^{\pm } &=&qX_{ab}^{\pm }X_{a}^{\pm }\quad \quad \quad
\quad \,\,\,X_{b}^{\pm }X_{ab}^{\pm }=q^{-1}X_{ab}^{\pm }X_{a}^{\pm }\quad
\,\,1\leq a<b\leq 3  \notag \\
X_{12}^{\pm }X_{13}^{\pm } &=&qX_{13}^{\pm }X_{12}^{\pm }\quad \quad \quad
\,\,\,\,X_{23}^{\pm }X_{13}^{\pm }=q^{-1}X_{13}^{\pm }X_{23}^{\pm }  \notag
\\
\left[ X_{2}^{\pm },X_{13}^{\pm }\right] &=&0\qquad \quad \quad \quad \quad
\,\,\left[ X_{2}^{\pm },X_{13}^{\mp }\right] =0\qquad  \notag \\
\left[ X_{12}^{+},X_{13}^{-}\right] &=&-q^{2\left( H_{1}+H_{2}\right)
}X_{3}^{-}\quad \left[ X_{12}^{-},X_{13}^{+}\right] =X_{3}^{+}q^{-2\left(
H_{1}+H_{2}\right) }  \notag \\
\qquad \left[ X_{23}^{+},X_{13}^{-}\right] &=&X_{1}^{-}\,q^{-2\left(
H_{2}+H_{3}\right) }\quad \left[ X_{23}^{-},X_{13}^{+}\right] =-q^{2\left(
H_{2}+H_{3}\right) }X_{1}^{+}\,  \notag \\
\left[ X_{12}^{\pm },X_{23}^{\pm }\right] &=&\lambda X_{2}^{\pm }X_{13}^{\pm
}\quad \quad \quad \,\left[ X_{12}^{\pm },X_{23}^{\mp }\right] =-\lambda
q^{\pm H_{2}}X_{1}^{\pm }X_{3}^{\mp }
\end{eqnarray}

\bigskip

where $\lambda =q-q^{-1}$.

Let $\mathcal{G}=su\left( 2,2\right) \cong so\left( 4,2\right) $ with
generators:

\bigskip

\begin{equation}
M_{AB}=-M_{BA},\,\,\,\,\,\,\, A\,,B=1,2,3,5,6,0,\,\,\eta _{AB}=diag\left(
-,-,-,-,+,+\right)
\end{equation}

\bigskip

which obey

\bigskip

\begin{equation}
\left[ M_{AB},M_{CD}\right] =i\left( \eta _{BC}M_{AD}-\eta _{CD}M_{BD}-\eta
_{BD}M_{AC}+\eta _{AD}M_{BC}\right) .
\end{equation}

\bigskip

Besides the ``physical'' generators $M_{AB}$ we shall also use the
``mathematical'' generators $Y_{AB}=-iM_{AB}$. Consider the Bruhat
decomposition: $\mathcal{G}=\mathcal{A}\oplus \mathcal{M}\oplus \widetilde{%
\mathcal{N}}\oplus \mathcal{N}$ (direct sum of vector spaces), where the
dilatation subalgebra $\mathcal{A}=so\left( 1,1\right) $ generated by $%
D=Y_{56}$ is a non-compact abelian subalgebra, the Lorentz subalgebra $%
\mathcal{M}=so(3,1)$ (a reductive Lie algebra) generated by $Y_{\mu \nu
}\,\left( \mu ,\upsilon =1,2,3,0\right) $ is the centralizer of $\mathcal{A}$
(mod $\mathcal{A}$), and the subalgebra of translations $\widetilde{\mathcal{%
N}}$ generated by $P_{\mu }=Y_{\mu 5}+Y_{\mu 6}$, the subalgebra of special
conformal transformations $\mathcal{N}$ generated by $K_{\mu }=Y_{\mu
5}-Y_{\mu 6}$, respectively, are nilpotent subalgebras forming the positive,
negative, respectively, root spaces of the root system ($\mathcal{G}$,$%
\mathcal{A}$).

Since $su\left( 2,2\right) $ is the conformal algebra of four dimensional
Minkowski spacetime we would like to deform it consistently with the
subalgebra structure relevant for the physical applications. The commutation
relations besides those for the Lorentz subalgebra are

\bigskip

\begin{eqnarray}
\left[ D,Y_{\mu \nu }\right] &=&0\qquad \left[ D,P_{\mu }\right] =0\qquad %
\left[ D,K_{\mu }\right] =-K_{\mu }  \notag \\
\left[ Y_{\mu \nu },P_{\lambda }\right] &=&\eta _{\nu \lambda }P_{\mu }-\eta
_{\mu \lambda }P_{\nu }\qquad \left[ Y_{\mu \nu },K_{\lambda }\right] =\eta
_{\nu \lambda }K_{\mu }-\eta _{\mu \lambda }K_{\nu }  \notag \\
\left[ P_{\mu },K_{\nu }\right] &=&2Y_{\mu \nu }+2\eta _{\mu \nu }D.
\end{eqnarray}

\bigskip

The algebra $\mathcal{P}_{\max }=\mathcal{M}\oplus \mathcal{A}\oplus 
\mathcal{N}$ (or equivalently $\widetilde{\mathcal{P}}_{\max }=\mathcal{M}%
\oplus \mathcal{A}\oplus \widetilde{\mathcal{N}}$ ) is the so called maximal
parabolic subalgebra of $\mathcal{G}$.

For the Lorentz algebra generators we have the following expressions

\bigskip

\begin{eqnarray}
H &=&-Y_{30}=\frac{1}{2}\left( H_{1}+H_{3}\right) \quad M^{\pm }=-iY_{31}\pm
iY_{10}=X_{1}^{\pm }+X_{3}^{\pm }  \notag \\
\widetilde{D} &=&-Y_{12}=\frac{i}{2}\left( H_{1}-H_{3}\right) \quad N^{\pm
}=-iY_{20}\pm iY_{23}=iX_{1}^{\pm }-iX_{3}^{\pm }.
\end{eqnarray}

\bigskip

For dilatation, translations and special conformal transformations we have

\bigskip

\begin{eqnarray}
D &=&\frac{1}{2}\left( H_{1}+H_{3}\right) +H_{2}  \notag \\
P_{0} &=&i\left( X_{13}^{+}+X_{+}^{2}\right) \qquad \qquad P_{1}=i\left(
X_{12}^{+}+X_{23}^{+}\right)  \notag \\
P_{2} &=&X_{12}^{+}-X_{23}^{+}\qquad \quad \qquad P_{3}=i\left(
X_{2}^{+}-X_{13}^{+}\right)  \notag \\
K_{0} &=&-i\left( X_{13}^{-}+X_{2}^{-}\right) \quad \,\,\,\, \quad
K_{1}=i\left( X_{12}^{-}+X_{23}^{-}\right)  \notag \\
K_{2} &=&X_{23}^{-}-X_{12}^{-}\qquad \quad \qquad K_{3}=i\left(
X_{2}^{-}-X_{13}^{-}\right) .
\end{eqnarray}

\bigskip

To derive the relations in $U_{q}\left( su\left( 2,2\right) \right) $ we use
equations (2.1) and (2.5).

The formulae for coproduct are:

\bigskip

\begin{eqnarray}
\Delta \left( H_{i}\right) &=&H_{i}\otimes 1+1\otimes H_{i}  \notag \\
\Delta \left( X_{\pm i}\right) &=&X_{\pm i}\otimes
q^{H_{i}/2}+q^{-H_{i}/2}\otimes X_{\pm i}.
\end{eqnarray}

\bigskip

The antipode and counit are defined as

\bigskip

\begin{eqnarray}
S\left( H_{i}\right) &=&-H_{i},  \notag \\
S\left( X_{+i}\right) &=&-qX_{+i},\qquad S\left( X_{-i}\right)
=-q^{-1}X_{-i},  \notag \\
\epsilon \left( H_{i}\right) &=&\epsilon \left( X_{+i}\right) =0.
\end{eqnarray}

\bigskip

\section{$q$-deformed conformal correlation functions}

\setcounter{equation}{0} First let us compute the $q$-deformed 2-point
conformal correlation function of scalar quasiprimary (qp) fields, with
canonical dimension $d_{1}$ and $d_{2}$, defined on the $q$-deformed
Minkowski spacetime\footnote{%
Up to Eq. (3.9) section 3 follows the paper \cite{dobrev4}.} \cite{dobrev4}:

\bigskip

\begin{eqnarray}
x_{\pm }v &=&q^{\pm 1}vx_{\pm },\quad \quad \quad \quad \qquad \,\,\, \qquad
x_{\pm }\overline{v}=q^{\pm 1}\overline{v}x_{\pm },  \notag \\
\lambda v\overline{v} &=&x_{+}x_{-}-x_{-}x_{+},\quad \,\, \quad \qquad
\qquad \overline{v}v=v\overline{v},  \notag \\
x_{\pm } &\equiv &x^{0}\pm x^{3}\quad \,\,\,\,\quad v\equiv
x^{1}-ix^{2}\quad \quad \overline{v}\equiv x^{1}+ix^{2}.
\end{eqnarray}

\bigskip

The $q$-Minkowsi length is

\bigskip

\begin{equation}
\mathcal{L}_{q}=x_{-}x_{+}-q^{-1}v\overline{v}.
\end{equation}

\bigskip

These qp-fields are reduced functions and can be written as formal power
series in the $q$-Minkowski coordinates:

\bigskip

\begin{eqnarray}
\phi &=&\phi \left( Y\right) =\phi \left( v,x_{-},x_{+},\overline{v}\right) 
\notag \\
&=&\sum_{j,n,l,m\in Z_{+}}\mu _{jnlm}\,\phi _{j\,\,n\,\,l\,\,m},  \notag \\
\phi _{jnlm} &=&v^{j}x_{-}^{n}x_{+}^{l}\overline{v}^{m}.
\end{eqnarray}

\bigskip

Next we introduce the following operators acting on the reduced functions as 
\begin{eqnarray}
\widehat{M}_{\kappa }\phi \left( Y\right) &=&\sum_{j,n,l,m\in Z_{+}}\mu
_{jnlm}\,\widehat{M}_{\kappa }\,\phi _{j\,\,n\,\,l\,\,m}  \notag \\
T_{\kappa }\phi \left( Y\right) &=&\sum_{j,n,l,m\in Z_{+}}\mu
_{jnlm}T_{\kappa }\phi _{j\,\,n\,\,l\,\,m}
\end{eqnarray}

\bigskip where $\kappa =\pm ,v,\overline{v}$ and the explicit action on $%
\phi _{j\,\,n\,\,l\,\,m}$ is defined by 
\begin{eqnarray}
\widehat{M}_{v}\phi _{j\,\,n\,\,l\,\,m} &=&\phi _{j+1\,\,n\,\,l\,\,m}  \notag
\\
\widehat{M}_{-}\phi _{j\,\,n\,\,l\,\,m} &=&\phi _{j\,\,n+1\,\,l\,\,m}  \notag
\\
\widehat{M}_{+}\phi _{j\,\,n\,\,l\,\,m} &=&\phi _{j\,\,n\,\,l+1\,\,m}  \notag
\\
\widehat{M}_{\overline{v}}\phi _{j\,\,n\,\,l\,\,m} &=&\phi
_{j\,\,n\,\,l\,\,m+1}  \notag \\
T_{v}\phi _{j\,\,n\,\,l\,\,m} &=&q^{j}\phi _{j\,\,n\,\,l\,\,m}  \notag \\
T_{-}\phi _{j\,\,n\,\,l\,\,m} &=&q^{n}\phi _{j\,\,n\,\,l\,\,m}  \notag \\
T_{+}\phi _{j\,\,n\,\,l\,\,m} &=&q^{l}\phi _{j\,\,n\,\,l\,\,m}  \notag \\
T_{\overline{v}}\phi _{j\,\,n\,\,l\,\,m} &=&q^{m}\phi _{j\,\,n\,\,l\,\,m}.
\end{eqnarray}

\bigskip

The $q$-difference operators are defined by

\bigskip

\begin{equation}
\widehat{D}_{\kappa }\phi =\frac{1}{\lambda }\widehat{M}_{\kappa
}^{-1}\left( T_{\kappa }-T_{\kappa }^{-1}\right) \phi .
\end{equation}

\bigskip

The representation action of $U_{q}\left( sl\left( 4\right) \right) $ on the
reduced functions $\phi \left( Y\right) $ of the representation space $%
C^{\Lambda }$, with the signature $\chi =\chi \left( \Lambda \right) =\left(
m_{1},m_{2},m_{3}\right) =\left( 1,1-d,1\right) $ and which corresponds to a
spinless \textquotedblleft scalar\textquotedblright\ field $\left[
d,j_{1},j_{2}\right] =\left[ d,0,0\right] $ is given by\footnote{%
The general case is given in Ref. \cite{dobrev2}} :

\bigskip

\begin{eqnarray}
\pi \left( k_{1}\right) \phi _{j\,\,n\,\,l\,\,m} &=&q^{\left( j-n+l-m\right)
/2}\phi _{j\,\,n\,\,l\,\,m},  \notag \\
\pi \left( k_{2}\right) \phi _{j\,\,n\,\,l\,\,m} &=&q^{n+\left( j+m+d\right)
/2}\phi _{j\,\,n\,\,l\,\,m},  \notag \\
\pi \left( k_{3}\right) \phi _{j\,\,n\,\,l\,\,m} &=&q^{\left(
-j-n+l+m\right) /2}\phi _{j\,\,n\,\,l\,\,m},  \notag \\
\pi \left( X_{+1}\right) \phi _{j\,\,n\,\,l\,\,m} &=&q^{-1+\left(
j-n-l+m\right) /2}\left[ n\right] _{q}\phi _{j+1\,\,n-1\,\,l\,\,m}  \notag \\
&&+q^{-1+\left( j-n+l-m\right) /2}\left[ m\right] _{q}\phi
_{j\,\,n\,\,l+1\,\,m-1},  \notag \\
\pi \left( X_{+2}\right) \phi _{j\,\,n\,\,l\,\,m} &=&q^{\left( -j+m\right)
/2}\left[ j+n+m+d\right] _{q}\phi _{j\,\,n+1\,\,l\,\,m}  \notag \\
&&+q^{d+\left( j+n+3m\right) /2}\left[ l\right] _{q}\phi
_{j+1\,\,n\,\,l-1\,\,m+1},  \notag \\
\pi \left( X_{+3}\right) \phi _{j\,\,n\,\,l\,\,m} &=&-q^{-1+\left(
j+n-l-m\right) /2}\left[ j\right] _{q}\phi _{j-1\,\,n\,\,l+1\,\,m}  \notag \\
&&-q^{-1+\left( 3j+n-3l-m\right) /2}\left[ n\right] _{q}\phi
_{j\,\,n-1\,\,l\,\,m+1},  \notag
\end{eqnarray}

\begin{eqnarray}
\pi \left( X_{-1}\right) \phi _{j\,\,n\,\,l\,\,m} &=&q^{2+\left(
-j+n-l+m\right) /2}\left[ j\right] _{q}\phi _{j-1\,\,n+1\,\,l\,\,m}  \notag
\\
&&+q^{2+\left( j-n-l+m\right) /2}\left[ l\right] _{q}\phi
_{j\,\,n\,\,l\,-1\,m+1},  \notag \\
\pi \left( X_{-2}\right) \phi _{j\,\,n\,\,l\,\,m} &=&-q^{\left( j-m\right)
/2}\left[ n\right] _{q}\phi _{j\,\,n-1\,\,l\,\,m},  \notag \\
\pi \left( X_{-3}\right) \phi _{j\,\,n\,\,l\,\,m} &=&-q^{\left(
-j-3n+l+3m\right) /2}\left[ l\right] _{q}\phi _{j+1\,\,n\,\,l-1\,\,m}  \notag
\\
&&-q^{\left( -j-n+l+m\right) /2}\left[ m\right] _{q}\phi
_{j\,\,n+1\,\,l\,\,m-1},
\end{eqnarray}

\bigskip

with $k_{i}=q^{H_{i}/2}$.

Now let us define $\mathcal{D}=q^{D}$, where $D$ is the dilatation generator
defined in Eq. (2.10). The representation of this generator on the reduced
functions $\phi $ is given by

\bigskip

\begin{eqnarray}
\pi \left( \mathcal{D}\right) \phi \left( Y\right) &=&\mu _{jnlm}\pi \left( 
\mathcal{D}\right) \phi _{j\,\,n\,\,l\,\,m}  \notag \\
\, &=&q^{d}\mu _{jnlm}q^{j+n+l+m}\phi _{j\,\,n\,\,l\,\,m}=q^{d}\phi \left(
qY\right) .
\end{eqnarray}

\bigskip

The coproduct for this operator is given by

\bigskip

\begin{equation}
\Delta \mathcal{D}=\mathcal{D}\otimes \mathcal{D}\text{.}
\end{equation}

\bigskip

Now let us calculate two point $q$-correlation functions by imposing that
they are invariant under the action of $U_{q}\left( sl\left( 4,\mathbb{C}
\right) \right) $. We denote the $q$-deformed correlation functions of $N$
quasiprimary fields as

\begin{equation}
\left\langle \phi _{1}\left( Y_{1}\right) ...\phi _{N}\left( Y_{N}\right)
\right\rangle _{q}=\,_{q}\left\langle 0\left\vert \phi _{d_{1}}\left(
Y_{1}\right) ...\phi _{d_{N}}\left( Y_{N}\right) \right\vert 0\right\rangle
_{q},
\end{equation}

\bigskip

where $\arrowvert0\rangle _{q}$ is a $U_{q}\left( sl\left( 4,\mathbb{C}%
\right) \right) $ invariant vacuum such that $\pi \left( \mathcal{D}\right) $%
$\arrowvert0\rangle _{q}=\arrowvert0\rangle _{q}$, $\pi \left( X_{+i}\right) %
\arrowvert0\rangle _{q}=0$ and also for $_{q}\langle 0\arrowvert$ . The
identities for the two-point correlation functions of two quasiprimary
fields of the conformal weights $d_{1}$, $d_{2}$ are

\bigskip

\begin{eqnarray}
\Delta \left( \pi \left( \mathcal{D}\right) \right) \left\langle \phi
_{1}\left( Y_{1}\right) \phi _{2}\left( Y_{2}\right) \right\rangle _{q}
&=&\left( \pi \left( \mathcal{D}\right) \otimes \pi \left( \mathcal{D}%
\right) \right) \left\langle \phi _{1}\left( Y_{1}\right) \phi _{2}\left(
Y_{2}\right) \right\rangle  \notag \\
&=&\left\langle \phi _{1}\left( Y_{1}\right) \phi _{2}\left( Y_{2}\right)
\right\rangle _{q}
\end{eqnarray}

\bigskip

and

\begin{eqnarray}
&&\Delta \left( \pi \left( X_{\pm i}\right) \right) \left\langle \phi
_{1}\left( Y_{1}\right) \phi _{2}\left( Y_{2}\right) \right\rangle _{q}= 
\notag \\
&&\left( \pi \left( X_{\pm i}\right) \otimes q^{\pi \left( H_{i}/2\right)
}+q^{-\pi \left( H_{i}/2\right) }\otimes \pi \left( X_{\pm i}\right) \right)
.  \notag \\
&&\left\langle \phi _{1}\left( Y_{1}\right) \phi _{2}\left( Y_{2}\right)
\right\rangle _{q}=0.
\end{eqnarray}

\bigskip

The $q$-correlation functions are covariant under dilatation, whereas the
remaining identities lead to six $q$-difference equations.

Let us first note that

\bigskip

\begin{eqnarray}
\phi _{j+1\,\,n-1\,\,l\,\,m} &=&q^{j}v\left( x_{-}\right) ^{-1}\,\,\phi
_{j\,\,n\,\,l\,\,m},\quad  \notag \\
\phi _{j\,\,n\,\,l+1\,\,m-1} &=&q^{m}\phi _{j\,\,n\,\,l\,\,m}\,\,x_{+}\left( 
\overline{v}\right) ^{-1},
\end{eqnarray}

\bigskip

and so on,

\begin{eqnarray}
q^{\pm j/2}\phi \left( v,x_{-},x_{+},\overline{v}\right) &=&\phi \left(
q^{\pm 1/2}v,x_{-},x_{+},\overline{v}\right) ,  \notag \\
q^{\pm n/2}\phi \left( v,x_{-},x_{+},\overline{v}\right) &=&\phi \left(
v,q^{\pm 1/2}x_{-},x_{+},\overline{v}\right) ,...
\end{eqnarray}

\bigskip

and

\begin{eqnarray}
\left[ n\right] _{q}\phi &=&\lambda ^{-1}\left( \phi \left( v,qx_{-},x_{+},%
\overline{v}\right) -\phi \left( v,q^{-1}x_{-},x_{+},\overline{v}\right)
\right)  \notag \\
&=& \widehat{D}_{-}\,\phi \left( v,x_{-},x_{+},\overline{v}\right) ,  \notag
\\
\left[ m\right] _{q}\phi &=&\lambda ^{-1}\left( \phi \left( v,x_{-},x_{+},q%
\overline{v}\right) -\phi \left( v,x_{-},x_{+},q^{-1}\overline{v}\right)
\right)  \notag \\
&=& \widehat{D}_{\overline{v}}\,\phi \left( v,x_{-},x_{+},\overline{v}%
\right) ,...
\end{eqnarray}

\bigskip

and so forth.

The first identity for $X_{+1}$ is given by:

\bigskip

\begin{eqnarray}
&&q^{j}v_{1}\left( x_{-1}\right) ^{-1}\,\,\langle \widehat{D}_{-}\,\phi
_{1}\left( q^{1/2}v_{1},q^{-1/2}x_{-1},q^{-1/2}x_{+1},q^{1/2}\overline{v}%
_{1}\right) .  \notag \\
&&\phi _{2}\left( q^{1/2}v_{2},q^{-1/2}x_{-2},q^{1/2}x_{+2},q^{-1/2}%
\overline{v}_{2}\right) \rangle _{q}  \notag \\
&&+q^{m}\langle \widehat{D}_{\overline{v}}\phi _{1}\left(
q^{1/2}v_{1},q^{-1/2}x_{-1},q^{1/2}x_{+1},q^{-1/2}\overline{v}_{1}\right)
x_{+1}\left( \overline{v}_{2}\right) ^{-1}.  \notag \\
&&\phi _{2}\left( q^{1/2}v_{2},q^{-1/2}x_{-2},q^{1/2}x_{+2},q^{-1/2}%
\overline{v}_{2}\right) \rangle _{q}  \notag \\
&&+q^{j}v_{2}\left( x_{-2}\right) ^{-1}\,\,\langle \phi _{1}\left(
q^{1/2}v_{1},q^{-1/2}x_{-1},q^{1/2}x_{+1},q^{-1/2}\overline{v}_{-1}\right) .
\notag \\
&&\widehat{D}_{-}\phi _{2}\left(
q^{1/2}v_{2},q^{-1/2}x_{-2},q^{-1/2}x_{+2},q^{1/2}\overline{v}_{2}\right)
\rangle _{q}  \notag \\
&&+q^{m}\langle \phi _{1}\left(
q^{1/2}v_{1},q^{-1/2}x_{-1},q^{1/2}x_{+1},q^{-1/2}\overline{v}_{1}\right)
x_{+2}\left( \overline{v}_{2}\right) ^{-1}.  \notag \\
&&\widehat{D}_{\overline{v}}\phi _{2}\left(
q^{1/2}v_{2},q^{-1/2}x_{-2},q^{1/2}x_{+2},q^{-1/2}\overline{v}_{1}\right)
\rangle _{q}=0,
\end{eqnarray}

\bigskip

and five other $q$-difference equations.

The solution of these $q$-difference equations exists when the conformal
dimensions $d_{1}$ and $d_{2}$ are equal: $d_{1}=d_{2}=d$ and is determined
uniquely up to a constant. Let us use the twistors $Y=Y ^{\mu }\sigma _{\mu
} $. More explicitly, the matrices $Y$ are given by:

\bigskip

\begin{equation}
Y=\left( 
\begin{array}{cc}
x_{0}+x_{3} & x_{0}-ix_{2} \\ 
x_{0}+ix_{2} & x_{0}-x_{3}%
\end{array}%
\right) =\left( 
\begin{array}{cc}
x_{+} & v \\ 
\overline{v} & x_{-}%
\end{array}%
\right) ,
\end{equation}

\bigskip

where $x_{+},x_{-},v,\overline{v}$ are $q$-deformed Minkowski coordinates
defined in Eq. (3.1).

It is easy to see that the quantum determinant\footnote{%
We use Manin's notation \cite{manin}.}

\bigskip

\begin{equation}
det_{q}Y=x_{-}x_{+}-q^{-1}v\overline{v}\qquad \text{and}\qquad det_{q}\left(
Y_{1}-Y_{2}\right) =det_{q}\,Y_{1}\left( I-Y_{1}^{-1}Y_{2}\right) ,
\end{equation}

\bigskip

where $I$ is a $2\times 2$ identity matrix. The $q$-deformed two-point
conformal correlation function reads

\bigskip

\begin{equation}
\left\langle \phi _{1}\left( Y_{1}\right) \phi _{2}\left( Y_{2}\right)
\right\rangle _{q}=C\left( q\right) \,\,\left( det_{q}\,Y_{1}\right)
^{-d}\,\,\,\,\,_{1}\varphi _{0}\left( d;q^{1-d/2}Y_{1}^{-1}Y_{2}\right) ,
\end{equation}

\bigskip

where $C\left( q \right) $ is a constant and where the quantum
hypergeometric function with matricial argument is given by:

\bigskip

\begin{equation}
\,_{1}\varphi _{0}\left( d;Y\right) =det_{q}\prod_{l=0}^{\infty }\left(
I-q^{l}Y\right) ^{-1}\left( I-q^{d+l}Y\right) .
\end{equation}

\bigskip

One easily see that the $q$-correlation function reduces to the undeformed
conformal correlation function because $\,_{1}\varphi _{0}\left( d;Y\right) $
becomes $\,_{1}F_{0}\left( d;Y\right) =det\left( I-Y\right) ^{-d}$ in the
limit $q\rightarrow 1$.

The identities for the $q$-deformed three-point conformal correlation
functions read

\bigskip

\begin{eqnarray}
&&(\pi \left( X_{i}\right) \otimes q^{\pi \left( -H_{i}/2\right) }\otimes
q^{\pi \left( -H_{i}/2\right) }+q^{\pi \left( -H_{i}/2\right) }\otimes \pi
\left( X_{i}\right) \otimes q^{\pi \left( -H_{i}/2\right) }  \notag \\
&&+q^{\pi \left( -H_{i}/2\right) }\otimes q^{\pi \left( -H_{i}/2\right)
}\otimes \pi \left( X_{i}\right) )\left\langle \phi _{1}\left( Y_{1}\right)
\phi _{2}\left( Y_{2}\right) \phi _{3}\left( Y_{3}\right) \right\rangle
_{q}=0,
\end{eqnarray}

\bigskip

\begin{equation}
\left( \pi \left( \mathcal{D}\right) \otimes \pi \left( \mathcal{D}\right)
\otimes \pi \left( \mathcal{D}\right) \right) \left\langle \phi _{1}\left(
Y_{1}\right) \phi _{2}\left( Y_{2}\right) \phi _{3}\left( Y_{3}\right)
\right\rangle _{q}=\left\langle \phi _{1}\left( Y_{1}\right) \phi _{2}\left(
Y_{2}\right) \phi _{3}\left( Y_{3}\right) \right\rangle _{q}.
\end{equation}

\bigskip

The solutions are given by

\bigskip

\begin{eqnarray}
&&\left\langle \phi _{1}\left( Y_{1}\right) \phi _{2}\left( Y_{2}\right)
\phi _{3}\left( Y_{3}\right) \right\rangle _{q}=C_{ijk}\,\,  \notag \\
&&\left( det_{q}Y_{1}\right) ^{-\gamma _{12}^{3}}\,\,_{1}\varphi _{0}\left(
\gamma _{12}^{3};q^{1-d_{1}/2}Y_{1}^{-1}Y_{2}\right) .  \notag \\
&&\left( det_{q}Y_{2}\right) ^{-\gamma _{23}^{1}}\,\,_{1}\varphi _{0}\left(
\gamma _{23}^{1};q^{1-d_{2}/2}Y_{2}^{-1}Y_{3}\right) .  \notag \\
&&\left( det_{q}Y_{1}\right) ^{-\gamma _{31}^{2}}\,\,_{1}\varphi _{0}\left(
\gamma _{31}^{2};q^{1+\frac{d_{2}-d_{1}}{2}}Y_{1}^{-1}Y_{3}\right)
\end{eqnarray}

\bigskip

where $\gamma _{ij}^{k}=\frac{d_{k}-d_{i}-d_{j}}{2}$ and $C_{ijk}$ are the
structure constants.

\bigskip

In a recent paper \cite{mesref2}, we studied the quantum gauge theory on the
quantum anti-de Sitter space $AdS^{q} _{5}$ and computed the quantum
metrics. Given these results, we can study the quantum analogue of the
celebrated AdS/CFT correspondence \cite{maldacena} and generalize the
methods used in \cite{hoffmann} to the $q$-deformed case.

\bigskip

\textbf{Acknowledgments}

I would like to thank V. K. Dobrev and J. Lukierski for sending their useful
comments. I am very grateful to W. Rühl for interesting discussions.

\end{document}